\documentclass[11pt]{article}

\usepackage{graphicx}
\usepackage{amsmath}
\usepackage{amsfonts}
\usepackage{amssymb}
\usepackage{verbatim}

\renewcommand{\arraystretch}{1.3}

\catcode`\@=11
\def\marginnote#1{}

\newcount\hour
\newcount\minute
\newtoks\amorpm
\hour=\time\divide\hour by60
\minute=\time{\multiply\hour by60 \global\advance\minute by-\hour}
\edef\standardtime{{\ifnum\hour<12 \global\amorpm={am}%
        \else\global\amorpm={pm}\advance\hour by-12 \fi
        \ifnum\hour=0 \hour=12 \fi
        \number\hour:\ifnum\minute<10 0\fi\number\minute\the\amorpm}}
\edef\militarytime{\number\hour:\ifnum\minute<10 0\fi\number\minute}

\def\draftlabel#1{{\@bsphack\if@filesw {\let\thepage\relax
      \xdef\@gtempa{\write\@auxout{\string
          \newlabel{#1}{{\@currentlabel}{\thepage}}}}}\@gtempa \if@nobreak
    \ifvmode\nobreak\fi\fi\fi\@esphack} \gdef\@eqnlabel{#1}}
    \def\@eqnlabel{}
\def\@vacuum{}
\def\draftmarginnote#1{\marginpar{\raggedright\scriptsize\tt#1}}

\def\draft{
  \oddsidemargin -.5truein
  \def\@oddfoot{\footnotesize \sl preliminary draft \hfil
    \rm\thepage\hfil\sl\today\quad\militarytime}
  \let\@evenfoot\@oddfoot \overfullrule 3pt
    \let\label=\draftlabel
    \let\marginnote=\draftmarginnote
  \def\@eqnnum{(\theequation)\rlap{\kern\marginparsep\tt\@eqnlabel}%
    \global\let\@eqnlabel\@vacuum}

  }

\makeatletter
\newdimen\normalarrayskip              
\newdimen\minarrayskip                 
\normalarrayskip\baselineskip
\minarrayskip\jot
\newif\ifold             \oldtrue            \def\new{\oldfalse}
\def\arraymode{\ifold\relax\else\displaystyle\fi} 
\def\eqnumphantom{\phantom{(\theequation)}}     
\def\@arrayskip{\ifold\baselineskip\z@\lineskip\z@
     \else
     \baselineskip\minarrayskip\lineskip2\minarrayskip\fi}
\def\@arrayclassz{\ifcase \@lastchclass \@acolampacol \or
\@ampacol \or \or \or \@addamp \or
   \@acolampacol \or \@firstampfalse \@acol \fi
\edef\@preamble{\@preamble
  \ifcase \@chnum
     \hfil$\relax\arraymode\@sharp$\hfil
     \or $\relax\arraymode\@sharp$\hfil
     \or \hfil$\relax\arraymode\@sharp$\fi}}
\def\@array[#1]#2{\setbox\@arstrutbox=\hbox{\vrule
     height\arraystretch \ht\strutbox
     depth\arraystretch \dp\strutbox
     width\z@}\@mkpream{#2}\edef\@preamble{\halign
\noexpand\@halignto
\bgroup \tabskip\z@ \@arstrut \@preamble \tabskip\z@ \cr}%
\let\@startpbox\@@startpbox \let\@endpbox\@@endpbox
  \if #1t\vtop \else \if#1b\vbox \else \vcenter \fi\fi
  \bgroup \let\par\relax
  \let\@sharp##\let\protect\relax
  \@arrayskip\@preamble}
\def\eqnarray{\stepcounter{equation}%
              \let\@currentlabel=\theequation
              \global\@eqnswtrue
              \global\@eqcnt\z@
              \tabskip\@centering
              \let\\=\@eqncr

 \halign to \displaywidth\bgroup
    \eqnumphantom\@eqnsel\hskip\@centering
    $\displaystyle \tabskip\z@ {##}$%
    \global\@eqcnt\@ne \hskip 2\arraycolsep
         $\displaystyle\arraymode{##}$\hfil
    \global\@eqcnt\tw@ \hskip 2\arraycolsep
         $\displaystyle\tabskip\z@{##}$\hfil
         \tabskip\@centering
    &{##}\tabskip\z@\cr}
\newfont{\hr}{msbm10}
\newfont{\ams}{msam10}

\hoffset= - 0.9in        

\def\beq{\begin{equation}}
\def\eeq{\end{equation}}
\def\ba{\beq\new\begin{array}{c}}
\def\ea{\end{array}\eeq}
\def\be{\ba}
\def\ee{\ea}

\def\N2{${\cal N}=2$}

\def\1N{${\cal N}=1$}
\def\4N{${\cal N}=4$}
\def\nn{\nonumber}

\newdimen\linethick  \linethick=0.4pt
\newdimen\hboxitspace    \hboxitspace=5pt
\newdimen\vboxitspace    \vboxitspace=5pt

\def\fr#1{%
\beq\new
\vcenter{
\hrule height\linethick
          \hbox{\vrule width\linethick
                \kern\hboxitspace
                \vbox{\kern\vboxitspace
                      \hbox{$\begin{array}{c}\displaystyle#1
         \end{array}$}%
                      \kern\vboxitspace}%
                \kern\hboxitspace
                \vrule width\linethick}%
          \hrule height\linethick}%
\eeq}

\renewcommand{\tt}[1][mer]{\hbox{\tiny{#1}}}

\newcommand{\Tr}{\mathop{\rm Tr}\nolimits}

\def\Tr{{\rm Tr}\,}

\def\rp{[p\,]}
\def\SS{{^*\!S}}

\title{
{\bf HOMFLY and superpolynomials for figure eight knot\ \ \ \
in all symmetric and antisymmetric representations}
\vspace{.5cm}}
\author{{\bf H.Itoyama}\footnote{ {\small {\it
Department of Mathematics and Physics,
Osaka City University} and {\it Osaka City University Advanced Mathematical
Institute (OCAMI), Osaka, Japan}};
itoyama@sci.osaka-cu.ac.jp}, \ {\bf A.Mironov}\footnote{ {\small {\it
Lebedev Physics Institute} and {\it ITEP, Moscow, Russia}};
mironov@itep.ru; mironov@lpi.ru}, \ {\bf A.Morozov}\thanks{{\small
{\it ITEP, Moscow, Russia}}; morozov@itep.ru}, \ {\bf
And.Morozov}\thanks{{\small {\it Moscow State University} and {\it ITEP,
Moscow, Russia}};
Andrey.Morozov@itep.ru}\date{ }}

\begin{document}

\setcounter{footnote}{3}

\setcounter{tocdepth}{3}

\maketitle

\vspace{-6.5cm}

\begin{center}
\hfill FIAN/TD-04/12\\
\hfill ITEP/TH-14/12\\
\hfill OCU-PHYS-364
\end{center}

\vspace{3.5cm}

\begin{abstract}
Explicit answer is given for the HOMFLY polynomial
of the figure eight knot $4_1$ in arbitrary symmetric
representation $R=[p\,]$.
It generalizes the old answers for $p=1$ and $2$
and the recently derived results for $p=3,4$,
which are fully consistent with the Ooguri-Vafa conjecture.
The answer can be considered as a quantization of
the ${\mathfrak{H}}_R = {\mathfrak{H}}_{[1]}^{|R|}$ identity for
the "special" polynomials (they define the leading
asymptotics of HOMFLY at $q=1$),
and arises in a form, convenient for
comparison with the representation of the Jones polynomials as sums of
dilogarithm ratios. In particular, we construct a difference equation
("non-commutative ${\cal A}$-polynomial") in the representation variable $p$.
Simple symmetry transformation provides also a formula for
arbitrary antisymmetric (fundamental) representation $R=[1^p\,]$,
which also passes some obvious checks.
Also straightforward is a deformation from HOMFLY
to superpolynomials.
Further generalizations seem possible to arbitrary
Young diagrams $R$, but these expressions are harder to test
because of the lack of alternative results, even partial.
\end{abstract}

\bigskip


\section{Introduction}

Recently a program was originated for a systematic study
of knot polynomials \cite{ACJK,HOMFLY,sups,DGR,AG} as functions of all their
parameters,
including the dependencies on representations and on knots.
This is a very important problem, given the relevance
of Chern-Simons theory \cite{CS} and its non-trivial deformations
for future development and extension
of our knowledge about conformal, topological and
integrable systems, matrix models,
instanton calculus and all kinds of dualities between them
implied by the gauge/string correspondence,
such as the AdS/CFT and AGT relations.
Recent years the progress was dramatically
slowed down by the lack of systematic description
of correlators in Chern-Simons theory,
comparable with the theory of conformal blocks,
Nekrasov functions, Virasoro constraints,
Sieberg-Witten equations and Dijkgraaf-Vafa phases.
The complication here is appearance of new parameters,
like knot topology and representation, which considerably
enlarge the space of observables and poses new
combinatorial problems.
The main ingredients of our approach is the search
for formulas, which explicitly depend on at least one parameter
(in addition to the usual $A,q,t$, which are rather
{\it arguments} of the knot polynomials),
and emphasize on the set of methods developed in
adjacent fields: character calculus, Ward identities
and integrability properties.

The four keystone choices, which have just inspired
a fast progress in this field are:
\begin{itemize}
\item Families of knots or representations instead of
isolated examples (a given knot and a given representation)
\item Braids instead of knots
\item Braid group (${\cal R}$-matrix) representation \cite{RT,inds}
instead of skein relations \cite{skein}
\item Character expansions instead of other
group theory methods (broadly applied to the problem
in the past decades).
\end{itemize}

This approach, though still a piece of art rather
than a fully deductive method, already proved its
power in the study of
\begin{itemize}
\item torus superpolynomials \cite{DMMSS,MMSS}
(dependence on the $\beta$-deformation \cite{betadefo}
parameter)
\item generic $m\leq 5$-strand braids in the
fundamental representation \cite{MMMkn2,MMShallHOMFLY}
(dependence on the knot)
\item extension of knot polynomials to $\tau$-function like
quantities \cite{MMMkn1} (introduction and dependence on
the time-variables)
\end{itemize}

The remaining direction, the dependence on representation
variable, was well studied in the "classical" period
of knot theory in the case of Jones polynomials,
but only partial results exist for anything else except for
a remarkable progress made by the Indian group \cite{inds,Rama00}
(see also \cite{Kaw}).
This direction is especially interesting because of
the hidden Ward identities, a small piece of which were
discovered in \cite{Apol},
and existing evidence \cite{DiFu,GS} of applicability of the
AMM/EO recursion \cite{AMM/EO}, which allows one to identify
the representation variables with (some discretized and
multiparametric!) version of the genus-expansion parameter.
Very recently there was a breakthrough in the two cases
(underlined are the free parameters):
\begin{itemize}
\item Arbitrary symmetric representation for the $\beta$-deformed
Jones superpolynomial for the trefoil  and other 2-strand knots
(all torus)
\cite{FGS}:
${\cal K}= [2,\underline{2k+1}]$, $R=\underline{\rp }$,
$A=t^2$
\item Symmetric representation for arbitrary
3-strand HOMFLY polynomial (not obligatory torus)
\cite{IMMM3s}: ${\cal K}=\underline{(a_1,b_1|a_2,b_2|\ldots)}$,
$R=[2]$, $t=q$
\end{itemize}

\bigskip

In this paper we add one more 1-parametric result:

\bigskip

$\bullet$ \framebox[15.5cm]{Arbitrary symmetric representation for the
HOMFLY polynomial of the figure eight knot}\ :

\smallskip

\hspace{.25cm} the simplest non-torus knot, which is 3-strand:
${\cal K} = 4_1 =  (1,-1|1,-1)$, $R=\underline{\rp }$,
$t=q$.

\noindent
In fact, this immediately implies also the answer for all
fully antisymmetric (i.e. all fundamental) representations, $R=[1^p]$.

\bigskip

The notation corresponds to the now standard parametrization
of extended knot superpolynomials \cite{MMMkn1}: ${\cal P}^{\cal
B}_R\{p_k|q,t\}$,
with $t=q^\beta$.
For $t=q$ (i.e. $\beta = 1$) this reduces to the extended HOMFLY polynomial
${\cal H}^{\cal B}_R\{p_k|q\}
= {\cal P}^{\cal B}_R\{p_k|q,t=q\}$,
which, if further reduced to a one-dimensional "topological locus" in
the infinite-dimensional space of time-variables,
\be
p_k = p_k^* = \frac{A^k-A^{-k}}{q^k-q^{-k}}
\label{tolo}
\ee
becomes the ordinary HOMFLY polynomial
$^*{\cal H}^{\cal K}_R\{A|\,q\} = {\cal P}^{\cal B}_R\{p_k^*|\,q,t=q\}$,
which does {\it not} depend on the braid representation ${\cal B}$ of the knot
${\cal K}$
and, after a further restriction $A=q^N$, possesses
a QFT representation as an average of the Wilson line in $3d$ Chern-Simons
theory:
\be
^*{\cal H}^{\cal K}_R\{A=q^N|\,q\} =
\left< \Tr\!_R P\exp\left(\oint_{\cal K} {\cal A}\right) \right>_{CS}
\ee
with the action $\frac{\kappa}{4\pi}\Tr\left({\cal A}d{\cal A} +
\frac{2}{3}{\cal A}^3\right)$,
the gauge group $SL(N)$ and $q = \exp\left(\frac{2\pi i}{\kappa+N}\right)$.

\smallskip

\smallskip

For a given number $m$ of strands, the braid is parameterized by a
semi-infinite matrix of integers
\be
{\cal B} = (a_{11},a_{12},\ldots,a_{1,m-1}|\, a_{21}, a_{22},\ldots,
a_{2,m-1}|\, a_{31}, \ldots)
\ee
(if we go along the braid, at first, the first and the second strands
intertwine $a_{11}$ times,
then, the second and the third intertwine $a_{21}$ times and so on).
In the $3$-strand case we also denote $a_{1i}=a_1$, $a_{2i}=b_i$.
For most such matrices the knots are either {\it composite} so that the
corresponding
HOMFLY polynomials are reducible, or they are actually not knots but links.
The HOMFLY polynomials are irreducible for the {\it prime}
knots, the first few (with $\sum_{ij} a_{ij} \leq 10$) listed in the Rolfsen
table at \cite{katlas}.

There are further important reductions of the HOMFLY polynomials:

$\bullet$ to Alexander polynomials: $A=1$

$\bullet$ to Jones polynomials: $A=q^2$

$\bullet$ to "special" polynomials: $q\to 1$.

\section{The main result}

Our claim in this paper is that
\be
\boxed{
^*{\cal H}^{4_1}_{\rp }(A|\,q) = \SS_{\rp }(A|\,q)\left(1+\
\sum_{k=1}^{p} \frac{\rp !}{[k]![p-k]!} \prod_{i=0}^{k-1}
\{Aq^{p+i}\}\{Aq^{i-1}\}\right)
}
\label{main}
\ee
where $\{x\} = x-x^{-1}$ and $[x] = \frac{\{q^x\}}{\{q\}} =
\frac{q^x-q^{-x}}{q-q^{-1}}$\
is the "quantum number" and the "figure eight knot" $4_1 = (1,-1|1,-1)$
is the first non-torus knot, which possesses a $3$-strand braid
representation.
Finally
\be
\SS_{\rp }(A|\,q) = S_{\rp }\{p^*_k\} = \prod_{i=1}^{p}
\frac{\{Aq^{i-1}\}}{\{q^{i}\}}
\ee
are the symmetric Schur functions $S_R(p_k)$
taken at the special point in the space of time variables (\ref{tolo}),
at the topological locus. At this point, the Schur function is equal to the
quantum dimension of the
representation $R$ and is equal to the value of the HOMFLY polynomial for the
unknot (trivial knot),
$^*{\cal H}^{\emptyset}_{\rp }(A|\,q) = \SS_{\rp }(A|\,q)$.
For $A=q^N$ this is just a quantum dimension of the symmetric representation
$R=\rp $
of the group $SL(N)$.

In particular, for the first three symmetric representations one explicitly
has:
\be
\frac{^*{\cal H}^{4_1}_{[0]}(A|\,q)}{\SS_{[0]}(A|\, q)} = 1, \\
\frac{^*{\cal H}^{4_1}_{[1]}(A|\,q)}{\SS_{[1]}(A|\, q)} =
A^2-q^2+1-q^{-2}+A^{-2} = 1 +
\{Aq\}\{Aq^{-1}\}, \\
\frac{^*{\cal H}^{4_1}_{[2]}(A|\,q)}{\SS_{[2]}(A|\, q)} =
q^4A^4-(q^6+q^4-q^2+q^{-2})\,A^2+(q^6-q^4+3-q^{-4}+q^{-6})\,-\\
-\,(q^2-q^{-2}+q^{-4}+q^{-6})\,A^{-2}+q^{-4}A^{-4} = 1 +
[2]\{Aq^2\}\{Aq^{-1}\}+\{Aq^3\}\{Aq^2\}\{A\}\{Aq^{-1}\}, \\
\frac{^*{\cal H}^{4_1}_{[3]}(A|\,q)}{\SS_{[3]}(A|\, q)} = 1 +
[3]\{Aq^3\}\{Aq^{-1}\}+[3]\{Aq^4\}\{Aq^3\}\{A\}\{Aq^{-1}\}
+\{Aq^5\}\{Aq^4\}\{Aq^3\}\{Aq\}\{A\}\{Aq^{-1}\}
\label{mainexa}
\ee

\section{The answer for arbitrary antisymmetric representation}

HOMFLY polynomials and their character expansions possess a
$Z_2$-symmetry
\be\label{symas}
A,\ q,\ \SS_R \ \ \longleftrightarrow \ A,\ -\frac{1}{q}\,,\ \SS_{R'}
\ee
where $R'$ is a transposition of the Young diagram $R$.
It is important here that $S_{R'}\{p_k\} = S_R\big\{(-)^{k-1}p_k\big\}$.
This symmetry allows to convert (\ref{main}) for the single-row
Young diagram $R=[p\,]$ into an answer for its transposed counterpart,
i.e. arbitrary single-column diagram $R=[1^p\,]=[\underbrace{1,1,\ldots,1}_{p\
{\rm times}}]$:
\vspace{-0.2cm}
\be
\boxed{
^*{\cal H}^{4_1}_{[1^p\,]}(A|\,q) = \SS_{[1^p\,]}(A|\,q)\left(1+\
\sum_{k=1}^{p} \frac{\rp !}{[k]![p-k]!} \prod_{j=0}^{k-1}
\{Aq^{-p-j}\}\{Aq^{-j+1}\}\right)
}
\label{mainas}
\ee
It is easy to check that all the sign factors disappear from this expression.
The relevant Schur functions at topological locus are given by
\be
\SS_{[1^p\,]}(A|\,q) = S_{[1^p\,]}\{p^*_k\} = \prod_{j=1}^{p}
\frac{\{Aq^{1-j}\}}{\{q^{j}\}}
\ee

The first few examples of (\ref{mainas}) are:
\be
\frac{^*{\cal H}^{4_1}_{[11]}(A|\,q)}{\SS_{[11]}(A|\, q)} =  1 +
[2]\{Aq^{-2}\}\{Aq\}+\{Aq^{-3}\}\{Aq^{-2}\}\{A\}\{Aq\}, \\
\!\!\!\!\!\!\!\!\!\!\!\!
\frac{^*{\cal H}^{4_1}_{[111]}(A|\,q)}{\SS_{[111]}(A|\, q)} = 1 +
[3]\{Aq^{-3}\}\{Aq^{1}\}+[3]\{Aq^{-4}\}\{Aq^{-3}\}\{A\}\{Aq\}
+\{Aq^{-5}\}\{Aq^{-4}\}\{Aq^{-3}\}\{Aq^{-1}\}\{A\}\{Aq\}
\label{mainexas}
\ee
It is easy to see that the r.h.s. of these expressions turn into unity
at $A=q^2$, $A=q^3$ respectively.
At the same time $\SS_{111}(A|q)$ also vanishes at $A=q^2$.

\section{Comments on the shape of (\ref{main})}

The answer (\ref{main}) is, first of all, a straightforward
generalization of the available examples (\ref{mainexa}) from
\cite{Rama00,IMMM3s}
(for $\rp =[2]$) and \cite{IMMM3hr} (for $\rp =[3],[4]$).

The desire to look for the special shape of the answer, which is the
key to generalization, comes from considering the "special" polynomials.
The special polynomial \cite{IMMMsp} is obtained from HOMFLY polynomials at $q\to
1$:
\be
{\mathfrak{H}}_R^{\cal K}(A) = \lim_{q\rightarrow 1} \frac{^*{\cal H}_R^{\cal
K}(A|\,q)}{^*S_R(A|\,q)}
\label{speHdef}
\ee
Note that the limit is taken with fixed $A$, and both the HOMFLY polynomial
$^*{\cal H}_R$
and the quantum dimension $^*S_R$ are singular behaving as
$(q-q^{-1})^{-|R|}$.
Here $|R|$ is the number of boxes in the Young diagram $R$,
in particular, $|\rp | = p$.
The conjectured property of "special" polynomials reads as \cite{DMMSS}
\be
\boxed{{\mathfrak{H}}^{\cal K}_R(A) = \Big({\mathfrak{H}}_{[1]}^{\cal
K}(A)\Big)^{|R|}}
\label{spepro}
\ee
and is presumably valid for arbitrary ${\cal K}$ and $R$.
It is an amusing "dual" of a somewhat similar conjecture
\be\boxed{
{\mathfrak{A}}_{R}^{\cal K}(q) = {\mathfrak{A}}_{[1]}^{\cal
K}\left(q^{|R|}\right)}
\label{alepro}
\ee
for the Alexander polynomial
\be
{\mathfrak{A}}_{R}^{\cal K}(q) = \lim_{A\rightarrow 1}
\frac{^*{\cal H}_R^{\cal K}(A|\,q)}{^*S_R(A|\,q)}
\ee
in any symmetric or antisymmetric representations\footnote{This is also correct,
at least, for the torus knots in all representations described by the
corner Young diagrams.}.

Since ${\mathfrak{H}}^{4_1}_{[1]} = 1+\{A\}^2$,
it follows from (\ref{spepro}) that
$\frac{^*{\cal H}_R^{4_1}(A|\,q)}{\SS_R(A|\,q)}$
is a quantization of
\be
{\mathfrak{H}}^{4_1}_R(A) = \Big(1+\{A\}^2\Big)^{|R|}
= \sum_{j=0}^{|R|} \frac{|R|!}{j!(|R|-j)!}\ \{A\}^{2j}
\label{sigmarel}
\ee
This explains the shape of eq.(\ref{main}).

\bigskip

Additional support for this kind of representation for the answer
comes from the Pochhammer (dilogarithm) decompositions of the Jones
polynomials.

\bigskip

The last piece of support is provided by the formula from \cite{Kaw} for
$^*H_{\rp }^{4_1}(A|\,q)$
at the peculiar point $A=q^N$, $q=\exp\left({\pi i\over N+p-1}\right)$,
which is easily reproduced from our (\ref{main}).
Indeed, the main formula of \cite{Kaw},
\be
\frac{^*H_{\rp }^{4_1}}{\SS_{\rp}}\left(A=q^N\Big|\,
q=e^{\frac{i\pi}{N+p-1}}\right) =
\left(1+\sum_{j=1}^p \prod_{i=0}^{j-1} \left(2\sin {(p-i)\pi\over
N+p-1}\right)^2\right)
\label{jaf}
\ee
is based on the same kind of decomposition,
only the items are full squares, which is, indeed, the case at this special
point and
one can check that (\ref{jaf}) follows from (\ref{main}).

\section{Evidence in favor of (\ref{main})\label{favor}}

We have seven pieces of evidence:
\begin{itemize}
\item Eq.(\ref{main}) reproduces (\ref{mainexa})
\item For $q\to 1$ eq.(\ref{main}) reproduces the conjectured
special polynomials
\item Eq.(\ref{main}) is consistent with the interesting
formula (\ref{jaf}) from \cite{Kaw}, describing the value of
$^*{\cal H}_{[p]}^{4_1}(A|q)$ at the one-dimensional locus
$q=e^{\frac{i\pi}{N+p-1}}$, $A=q^N = -e^{\frac{i\pi (1-p)}{N+p-1}}$
\item For $A=q^2$   eq.(\ref{main}) reproduces
the known answers for the Jones polynomials, moreover, directly
in their most advanced form in terms of the dilogarithm ratios (products of the Pochhammer symbols)
\item For $A=1$ eq.(\ref{main}) reproduces
the Alexander polynomial,
which depends on the representation $R$ through (\ref{alepro})
\item According to (\ref{mainas}) the antisymmetric HOMFLY polynomial
$^*{\cal H}_{[1^p]}^{4_1}(A|\,q)$ vanishes for $A=q^N$
with $N<p$, i.e. whenever $p$ exceeds the rank of the group
by two, and turns its ratio to the unknot into unity for $N=p$.
Since (\ref{mainas}) and (\ref{main}) are related by the symmetry transform
(\ref{symas}), this
necessary property simultaneously checks (\ref{main})
\item Eq.(\ref{main}) is consistent with the Ooguri-Vafa
conjecture
\end{itemize}

\noindent
The first two points in this list are not independent checks,
since they were used to write (\ref{main}) down.
However, the last five are.
In what follows we briefly comment on the first and the last
stories in this list. Other points do not seem to require additional
explanations.

\section{On the derivation of (\ref{mainexa})}

The key source of intuition for the quantization rule
$(\ref{sigmarel})\ \longrightarrow\ (\ref{main})$
are explicit formulas (\ref{mainexa}).
They are obtained from the character expansion of
the extended HOMFLY polynomials for a generic
$3$-strand braid \cite{MMMkn1,MMMkn2,IMMM3s,IMMM3hr}:
\be
{\cal H}^{(a_1,b_1|a_2,b_2|\ldots)}_{R}\{p_k|q\}
= \sum_{Q\in R^{\otimes m}} S_Q\{p_k\}\cdot
\Tr_{N_{RQ}\otimes N_{RQ}} \Big\{\hat{\cal R}_{RQ}^{a_1}\hat U
\hat{\cal R}_{RQ}^{b_1}\hat U_{RQ}^\dagger
\hat{\cal R}_{RQ}^{a_2}\hat U_{RQ}
\hat{\cal R}_{RQ}^{b_2}\hat U_{RQ}^\dagger \ldots\Big\}
\label{3expan}
\ee
at the topological locus (\ref{tolo}).
In this particular case $m=3$, but equally simple
and general expressions can be written for arbitrary $m$,
see \cite{MMMkn1,MMMkn2}. The figure eight knot is characterized by
$a_1=a_2=1$, $b_1=b_2=-1$.

The sum here goes over the Young diagrams $Q\in R^{\otimes m}$,
which appear $N_{RQ}$ times in the decomposition of the $m$-the
power of representation $R$ and have the size $|Q|=m|R|$.
The Schur functions $S_Q\{p_k\}$ are well known for all $Q$.
The multiplicities $N_{RQ}$ define the sizes of the
quantum ${\cal R}$-matrices, projected onto the space of
the intertwining operators, we call them $\hat{\cal R}$
and they are diagonal matrices of the form
\be
\hat{\cal R} = {\rm diag}\Big(\, \epsilon_i\, q^{\varkappa_i}\
\Big|\  i = 1,\ldots, N_{RQ}\Big)
\ee
with certain exponents $\varkappa_i$ and signs $\epsilon_i=\pm$,
depending on the choice of $R$ and $Q\in R^{\otimes m}$.
The mixing matrices $\hat U_{RQ}$ describe the transition
between bases in the $N_{RQ}$-dimensional space of
intertwiners $R^{\otimes m} \longrightarrow Q$,  which
diagonalize $\hat{\cal R}$-matrices, associated with
the crossings of the strands $12$ and $23$ in the braid
(for $m>3$ there will be  $m-2$ different mixing matrices
for given $R$ and $Q$).
In practice, for $m=3$ these matrices depend only on $N_{RQ}$
and they are concrete functions of
the parameters $\{\epsilon_i,\varkappa_i\}$ in the corresponding
$\hat{\cal R}_{RQ}$.
These functions are explicitly found in \cite{IMMM3hr}.\footnote{
For $N_{RQ}\leq 3$ this result has been long ago obtained by a
slightly different method in the beautiful paper \cite{Rama00}.
Some other particular cases can be obtained using
the formulas for the $SU_q(2)$ Racah coefficients, \cite{Kl}.
This idea goes back to \cite{AGS} and was successfully developed in \cite{inds}.
}
Therefore

{\footnotesize
\be
^*{\cal H}_{[3]}^{4_1}(A| q) =
\SS_{[9]}(A| q)+ \Big(q^{12}-2q^{6}+1-2q^{-6}+q^{-12}\Big) \SS_{[81]}(A| q) +
\nn \\ \!\!\!\!\!\!\!\!\!\!\!\!\!\!\!
+\Big(q^{20}\!\!-2q^{16}\!\!-2q^{14}\!\!+q^{12}\!\!+4q^{10}\!\!+q^{8}\!-4q^{
6}\!-4q^{4}\!+2q^{2}+6
+2q^{-2}\!\!-4q^{-4}\!\!-4q^{-6}\!\!+q^{-8}\!\!+4q^{-10}\!\!+q^{-12}\!\!-2q^
{-14}\!\!-2q^{-16}
\!\! +q^{-20}\Big)\SS_{[72]}(A| q)+\nn\\
+\SS_{[711]}(A| q)+ \underline{\alpha_{[63]}\SS_{[63]}(A| q)}
+ \Big(q^{8}-2q^{4}+1-2q^{-4}+q^{-8}\Big)\SS_{[621]}(A| q)
+\Big(q^{8}-2q^{4}+1-2q^{-4}+q^{-8}\Big)\SS_{[54]}(A| q)+
\nn \\
+\Big(q^{12}-2q^{10}-q^{8}+4q^{6}-3q^{4}-2q^{2}+6
-2q^{-2}-3q^{-4}+4q^{-6}-q^{-8}-2q^{-10}+q^{-12}\Big)\SS_{[531]}(A| q) +\nn\\
+ \SS_{[522]}(A|\ q)
+\SS_{[441]}(A|  q)+ \Big(q^{4}-2q^{2}+1-2q^{-2}+q^{-4}\Big)\SS_{[432]}(A|
q)+
\SS_{[333]}(A|  q)
\ee
}
\!\!where $\alpha_{[63]}$ in the underlined term
is given by a trace of the $4\times 4$ matrix and is equal to
{\footnotesize
\be
\alpha_{[63]} =
q^{24}-2q^{22}-q^{20}+2q^{18}+3q^{16}-2q^{14}-6q^{12}
+4q^{10}+8q^8-2q^6-12q^4+2q^2+11 + \nn \\
+2q^{-2}-12q^{-4}-2q^{-6}+8q^{-8}+4q^{-10}-6q^{-12}
-2q^{-14}+3q^{-16}+2q^{-18}-q^{-20}-2q^{-22}+q^{-24}
\ee
}
\!\!This formula reproduces the known answers for the colored Jones polynomial
($A=q^2$) from \cite{katlas}\footnote{
The answer is also consistent with the
Alexander polynomial ($A=1$) in \cite{katlas}, but $\SS_{[63]}(A|\,q)$
vanishes at $A=1$
and, therefore, this comparison is insensitive to the value
$\alpha_{[63]}$. Similarly, comparison with the "special" polynomial
${\mathfrak{H}}_{[3]}^{4_1}(A)$ allows one only to check the value of
$\alpha_{[63]}$ at $q=1$.}
and after the substitution of explicit expressions for $\SS_Q(A|q)$
it  coincides with (\ref{main}).

For $R=[4]$ there will be two items with $4\times 4$ matrices and one item
with a $5\times 5$ matrix:

{\footnotesize
\be
^*{\cal H}_{[4]}^{4_1}(A|\, q) =
\SS_{[12]}(A|\,q)\ +  \Big(q^{16}-2q^{8}+1-2q^{-8}+q^{-16}\Big)\SS_{[11,
1]}(A|\,q)\
+ \Big(q^{28}\!\!-2q^{22}\!\!-2q^{20}\!\!+q^{16}\!\!+4q^{14}\!\!+q^{12}-\nn \\
-4q^{8}\!-4q^{6}\!+2q^{2}\!
+6+2q^{-2}\!\!-4q^{-6}\!\!-4q^{-8}\!\!+q^{-12}\!\!+4q^{-14}\!\!+q^{-16}\!\!-2q^{-20}\!\!
-2q^{-22}\!\!+q^{-28}\Big)\SS_{[10, 2]}(A|\,q)\ +\nn\\
+  \SS_{[10, 1, 1]}(A|\,q)\ +  \underline{\alpha_{[9,3]}\SS_{[9, 3]}(A|\,q)}\
+  \Big(q^{12}-2q^{6}+1-2q^{-6}+q^{-12}\Big)\SS_{[9, 2, 1]}(A|\,q)\ +
\underline{\underline{\alpha_{[8,4]}\SS_{[8, 4]}(A|\,q)}}\ +\nn\\
+
\Big(q^{20}\!\!-2q^{16}\!\!-2q^{14}\!\!+q^{12}\!\!+4q^{10}\!\!+q^{8}\!-4q^{6}\!-4q^{4}\!+2q^{2}\!+6
+2q^{-2}\!\!-4q^{-4}\!\!-4q^{-6}\!\!+q^{-8}\!\!+4q^{-10}\!\!+q^{-12}-\nn\\
-2q^{-14}\!\!-2q^{-16}\!\!
+q^{-20}\Big)\Big( \SS_{[8, 3, 1]}(A|\,q)\ + \SS_{[7,5]}(A|\,q)\Big)
+\SS_{[8,2,2]}(A|\,q)\ +  \underline{\alpha_{[7,4,1]}\SS_{[7, 4, 1]}(A|\,q)}\
+
\nn\\+
\Big(q^{8}-2q^{4}+1-2q^{-4}+q^{-8}\Big)\Big(\SS_{[7, 3, 2]}(A|\,q)\  +
\SS_{[6,
5, 1]}(A|\,q)\Big)
+  \SS_{[6, 6]}(A|\,q)\  + \nn\\
+\Big(q^{12}-2q^{10}-q^{8}+4q^{6}-3q^{4}-2q^{2}+6-2q^{-2}-3q^{-4}+4q^{-6}-q^{-8}-2q^{-10}+q^{-12}\Big)
\SS_{[6, 4, 2]}(A|\,q)\ + \nn \\
+  \SS_{[6, 3, 3]}(A|\,q) +  \SS_{[5, 5, 2]}(A|\,q)\ +
\Big(q^{4}-2q^{2}+1-2q^{-2}+q^{-4}\Big)\SS_{[5, 4, 3]}(A|\,q)\ +  \SS_{[4, 4,
4]}(A|\,q)
\ee
}
\!\!with
{\footnotesize
\be
\alpha_{[9,3]}\
=q^{36}-2q^{32}-2q^{30}-q^{28}+4q^{26}+5q^{24}+2q^{22}-5q^{20}-10q^{18}-3q^{16}+8q^{14}+14q^{12}+
\nn \\
+6q^{10}-9q^{8}-18q^{6}-9q^{4}+10q^{2}+19
+10q^{-2}-9q^{-4}-18q^{-6}-9q^{-8}+6q^{-10}+\nn \\
+14q^{-12}+8q^{-14}-3q^{-16}-10q^{-18}-5q^{-20}+2q^{-22}
+5q^{-24}+4q^{-26}-q^{-28}-2q^{-30}-2q^{-32}+q^{-36}
 \nn \\ \nn \\
\alpha_{[8,4]}\
=q^{40}-2q^{38}-q^{36}+2q^{34}+q^{32}+4q^{30}-6q^{28}-4q^{26}+2q^{24}+4q^{22}+13q^{20}
-10q^{18}-15q^{16}-4q^{14}+\nn \\
+10q^{12}+28q^{10}-6q^{8}-24q^{6}-18q^{4}+6q^{2}+37+6q^{-2}-18q^{-4}-24q^{-6}-6q^{-8}+28q^{-10}+10q^{-12}
-4q^{-14}-\nn \\
-15q^{-16}-10q^{-18}+13q^{-20}+4q^{-22}+2q^{-24}
-4q^{-26}-6q^{-28}+4q^{-30}+q^{-32}+2q^{-34}-q^{-36}-2q^{-38}+q^{-40}
 \nn \\ \nn \\
\alpha_{[7,4,1]}\ =
q^{24}-2q^{22}-q^{20}+2q^{18}+3q^{16}-2q^{14}-6q^{12}+4q^{10}+8q^{8}-2q^{6}
-12q^{4}+2q^{2}+11+\nn \\
+2q^{-2}-12q^{-4}-2q^{-6}+8q^{-8}+4q^{-10}-6q^{-12}-2q^{-14}+3q^{-16}+2q^{-18}-q^{-20}-2q^{-22}+q^{-24}
 \nn \\ \nn \\
\ee
}
\!\!
Again, this expression reproduces the known Jones and Alexander polynomials
and again, Alexander is not sensitive to concrete values of the three $\alpha$-parameters.
Remarkably, these huge formulas, the best what can be directly deduced
from the braid calculus,
can be encapsulated into a simple expression (\ref{main})
on the topological locus (\ref{tolo}).

\bigskip

To conclude this section, we note that the
character expansion formulas like (\ref{3expan})
are very different in structure from the expansions like (\ref{main}),
implied by the quantization of the "special" polynomial
basic identity (\ref{sigmarel}), which is, in turn,
closely related to the Pochhammer decompositions of \cite{Poch}.
It will be extremely important to understand the relation
(duality?) between such decompositions in full generality.

\section{Check of the Ooguri-Vafa conjecture}

The HOMFLY polynomials are rather huge.
It is also possible to express them through somewhat
different Ooguri-Vafa (OV) polynomials \cite{OV}.
At least, in the case of the
figure eight knot
the OV polynomials are partly factorizable and, hence, look much simpler than
the original HOMFLY polynomials.
The relation between the two involves a relation
between the disconnected and connected correlators and is given by comparing
the generating functions of all
representations
\be
\boxed{\log\left(\sum_R {{\cal H}}_R^{\cal K}\big\{p_k|\,q\big\}S_R\{\bar
p_k\}\right)=
\sum_n \sum_R\ {1\over n}\ {{f}}_R^{\cal
K}\big\{p^{(n)}_k\big|\,q^n\big\}S_R\big\{\bar p^{(n)}_k\big\}}
\ee
and at the topological locus (\ref{tolo}) this reduces to
\be
\log\left(\sum_R {^*{\cal H}}_R^{\cal K}(A|\,q)S_R\{\bar
p_k\}\right)=
\sum_n \sum_R\ {1\over n}\ {^*\!{f}}_R^{\cal K}(A^n|q^n)S_R\big\{\bar
p^{(n)}_k\big\}
\ee
Here $\{\bar p_k\}$ is an additional set of arbitrary
time variables, and $\bar p_k^{(n)}\equiv
\bar p_{nk}$ is their Adams transform.
The OV conjecture claims \cite{OV} that, like the ratios $^*{\cal
H}_R/{^*\!S}_R$,
the ratio $^*\!f_R/\SS_{[1]}$
is always a polynomial with
integer coefficients. This means that $^*\!f_R$ is always proportional to
$\SS_{[1]}$, i.e. is less singular then $^*{\cal H}_R$, in the
("special-polynomial") limit of $q\rightarrow 1$ with $A$ fixed.

The first few OV polynomials are
\be
f_{[1]}={\cal H}_{[1]}\\
f_{[2]}={\cal H}_{[2]}-{1\over 2}{\cal H}_{[1]}^2-
{1\over 2}{\cal H}_{[1]}^{(2)}\\
f_{[3]}={\cal H}_{[3]}-{\cal
H}_{[1]}{\cal H}_{[2]}+
{1\over 3}{\cal H}_{[1]}^3-
{1\over 3}{\cal H}_{[1]}^{(3)}\\
f_{[4]}={\cal H}_{[4]}-{\cal
H}_{[1]}{\cal H}_{[3]}-
{1\over 2}{\cal H}_{[2]}^2+
{\cal H}_{[1]}^2{\cal H}_{[2]}-{1\over 4}{\cal
H}_{[1]}^4
+{1\over 2}{\cal H}_{[2]}^{(2)}+{1\over 4}\left({\cal
H}_{[1]}^{(2)}\right)^2
\ee
where
${\cal H}_R^{(n)}\{p_k|\,q\})\equiv {\cal H}_R\{p^{(n)}_k|\,q^n\}$
and
$^*{\cal H}_R^{(n)}(A|\,q)\equiv \phantom{.}^*{\cal H}_R(A^n|\,q^n)$.

In the examples (\ref{mainexa}) one has:
\be
\frac{^*\!{f}^{4_1}_{[0]}(A|\,q)}{\SS_{[1]}(A|\, q)} = 0\\
\frac{^*\!{f}^{4_1}_{[1]}(A|\,q)}{\SS_{[1]}(A|\, q)}
=A^2-(q^2-1+q^{-2})+A^{-2}
\\
\frac{^*\!{f}^{4_1}_{[2]}(A|\,q)}{\SS_{[1]}(A|\, q)} =
\{A\}\{A/q\}\{Aq^2\}\{A^2q^2\} , \\
\frac{^*\!{f}^{4_1}_{[3]}(A|\,q)}{\SS_{[1]}(A|\, q)}
=\{A\}\{A/q\}\{Aq\}\{Aq^2\}
\Big((q^{8}+q^{4}+q^{2})A^4-(q^{6}-q^{4}-1)A^2-\\-(q^{2}-2
+q^{-2})-(q^{-6}-q^{-4}-1)A^{-2}+(q^{-8}+q^{-4}+q^{-2})A^{-4}
\Big)   \\
\frac{^*\!{f}^{4_1}_{[4]}(A|\,q)}{\SS_{[1]}(A|\, q)} =
\{A\}\{A/q\}\{Aq\}\{Aq^2\}
\Big(
(q^{19}+q^{15}+q^{13}+2q^{11}+q^{9}+2q^{7}+q^{5}+q^{3})A^{7}- \\
-\,(q^{17}+q^{15}+q^{13}+2q^{11}+2q^{9}+2q^{7}+q^{5})A^{5}+
\,(q^{13}+q^{9}-q^{5}-q^{3}-2q^{-1})A^{3}+\\
+\,(q^{7}+q^{5}+q^{3}+2q-q^{-5})A
+(q^{-7}+q^{-5}+q^{-3}+2q^{-1}-q^{5})A^{-1}+\\
+\,(q^{-13}+q^{-9}-q^{-5}-q^{-3}-2q^{1})A^{-3}
-\,(q^{-17}+q^{-15}+q^{-13}+2q^{-11}+2q^{-9}+2q^{-7}+q^{-5})A^{-5}+\\+
\,(q^{-19}+q^{-15}+q^{-13}+2q^{-11}+q^{-9}+2q^{-7}+q^{-5}+q^{-3})A^{-7}
\Big)\\
\ldots
\label{mainexaF}
\ee
For these polynomials the Ooguri-Vafa conjecture is obviously true.
It would be interesting to check it in full generality
(for arbitrary $\rp $) for the answer (\ref{main}).

In the special limit of $q\to 1$, the special OV polynomials
${\mathfrak{f}}_{R}(A)\equiv\lim_{q\rightarrow 1}{^*\!f_R(A|\,q)\over
\SS_{[1]}(A|\,q)}$
are
\be
{\mathfrak{f}}_{[0]}(A) = 1, \\
{\mathfrak{f}}_{[1]}(A) =  1+\{A\}^2, \\
{\mathfrak{f}}_{[2]}(A) = \{A\}^3\{A^2\}   , \\
{\mathfrak{f}}_{[3]}(A) = \{A\}^4
\Big(3A^4+A^2+A^{-2}+3A^{-4}\Big), \\
{\mathfrak{f}}_{[4]}(A) = 2\{A\}^3\{A^2\}\Big(
5A^{6}-10A^{4}+9A^2-7+9A^{-2}-10A^{-4}+5A^{-6}
\Big), \\
\ldots
\label{mainexaFs}
\ee
These polynomials non-trivially depend on representation,
and can not be expressed through the special
polynomials ${\mathfrak{H}}_R(A)$: higher order corrections in $1-q$
also contribute to ${\mathfrak{f}}_{[R]}(A)$,
see \cite{IMMMsp}.

Likewise one can easily obtain from (\ref{mainas})
the OV polynomials for
the antisymmetric representations.

\section{Colored superpolynomial for the figure eight knot}

Our main formula (\ref{main}) has, in fact, a very suggestive form,
calling for various generalizations.
However, for this purpose on should interpret it in a proper way.
As it is, (\ref{main}) looks like a sum over either
the sub-diagrams of the Young diagram $R$, or simply the
subsets of boxes in $R$.
It turns our that the adequate one is the second representation.
As we demonstrate in this section it provides an immediate suggestion
for the {\it super}polynomial ($\beta$-) deformation of (\ref{main}).

$\beta$-deformation implies that the shifts along the
horizontal and vertical axes in Young diagram produce
the $q$ and $t^{-1}$, rather than $q$ and $q^{-1}$ factors
in all the formulas. However, there is no immediate rule for
the substitution (refinement) of the binomial coefficients.
The $\beta$-deformation would be much simpler if sum over
boxes would contain only {\it unit} coefficients.
This means that in order to be straightforwardly deformed,
(\ref{main}) should first be rewritten in such a way.

This is, indeed, possible.   To begin with,
\be
[p]\{Aq^{p+i}\}\{Aa^{i-1}\}
= \sum_{i=1}^{p} \{Aq^{2(p-i)+1}\}\{Aq^{-1}\}
\ee
i.e. this contribution can be immediately rewritten
as a sum over boxes with unit coefficients.
This implies that with every box with coordinates $(i,1)\in [p]$
one associates a product $Z_i(A)=\{Aq^{2(p-i)+1}\}\{Aq^{-1}\}$.
Next,
\be
 \frac{[p][p-1]}{[2]}\{Aq^{p+i+1}\}\{Aq^{p+i}\}\{A\}\{Aq^{-1}\} = \\
= \sum_{1\leq i< i'\leq p} \{ Aq^{2(p-i)+1}\}\{Aq^{-1}\}
\{Aq^{2(p-i')+2}\}\{A\} = \sum_{1\leq i< i'\leq p} Z_i(A)Z_{i'}(Aq)
\ee i.e. the argument of the second factor $Z_{i'}$ is shifted by
the factor of $q$. In general (\ref{main}) can be rewritten as
\be\label{amain} \!\!\!\!\!\!\!\!\! \frac{^*{\cal H}^{4_1}_{\rp
}(A|\,q)}{ \SS_{\rp }(A|\,q)} \ =\
\sum_{k=0}^p\frac{[|R|]!}{[k]![|R|-k]!}\prod_{i=1}^{k}  Z_i(A) =
\sum_{k=0}^p\ \sum_{1\leq i_1<\ldots < i_k\leq
p}\!\!\!\!\!\!\!\!\! Z_{i_1}(A)Z_{i_2}(Aq)Z_{i_3}(Aq^2)\ldots
Z_{i_k}(Aq^{k-1}) \ee

This formula already admits a straightforward $\beta$-deformation:
it is enough to change
$Z_i(A)=\{Aq^{2(p-i)+1}\}\{Aq^{-1}\}$ for
${\mathfrak Z}_i(A)=\{Aq^{2(p-i)+1}\}\{At^{-1}\}$, and
\be
\boxed{
\frac{^*{\cal P}^{4_1}_{\rp }(A|\,q,t)}{ ^*\!M_{\rp }(A|\,q,t)}\ \ =\
 \sum_{k=0}^p\ \ \sum_{1\leq i_1<\ldots < i_k\leq p}\!\!\!\!\!
{\mathfrak Z}_{i_1}(A){\mathfrak Z}_{i_2}(Aq){\mathfrak Z}_{i_3}(Aq^2)\ldots {\mathfrak Z}_{i_k}(Aq^{k-1})
}
\label{mainP}
\ee
is the superpolynomial for knot $4_1$ in arbitrary symmetric representation $[p\,]$.
(Denominator at the l.h.s. is the value of the MacDonald polynomial at topological locus,
which is the standard choice for the superpolynomial unknot,
see \cite{DMMSS} and references therein).
The evidence in favor of this conjecture involves the two facts:

$\bullet$
For $\ t=q\ $ eq.(\ref{mainP}) reproduces our HOMFLY polynomial (\ref{main}).

$\bullet$
After the standard change of variables \cite{AG,DMMSS},
\be
t={\bf q},\ \ \ \ \ \ \ q=-{\bf q}{\bf t},\ \ \ \ \ \ \ \ \ \ \
A = {\bf a}\sqrt{-{\bf t}}
\label{newv}
\ee
the superpolynomial (\ref{mainP}) is, indeed, a polynomial in all its variables
${\bf a}^{\pm 1},\ {\bf q}^{\pm 1},\ {\bf t}^{\pm 1}\ $ with all positive coefficients,
simply because each factor ${\cal Z}_i(Aq^s)$ acquires this positivity property
in the new variables:
\be
{\mathfrak Z}_i(Aq^s) =\frac{\left(1-A^2q^{4(p-i)+2+2s}\right)\left(t^2-A^2q^{2s}\right)}
{(A^2tq)\cdot q^{2(p-i+s)}}= \frac{\left(1+{\bf a}^2{\bf t}({\bf qt})^{4(p-i)+2+2s}\right)
\left({\bf q}^2+{\bf a}^2{\bf t}({\bf qt})^{2s}\right)}
{ {\bf a}^2\cdot ({\bf qt})^{2(p-i+s+1)}}
\ee
The symmetry (\ref{symas}) has an obvious  generalization to the $\beta$-deformed
situation (see also \cite{FGS,GMS}), just
\be\label{symsp}
(q,t) \longrightarrow (-t^{-1},-q^{-1})\ \ \ \ \ \ \ \ \ \ \ \ \
\hbox{or} \ \ \ \ \ \ \ \ \ \ \ \ \ \ \ ({\bf q},{\bf t}) \longrightarrow (1/{\bf qt},{\bf t})
\ee
which allows one to convert (\ref{mainP}) into a similar formula for
the superpolynomials in the fundamental representations:
\be
\boxed{
\frac{^*{\cal P}^{4_1}_{[1^p\,]}(A|\,q,t)}{ ^*\!M_{[1^p\,]}(A|\,q,t)}\ \ =\
\sum_{k=0}^p\ \ \sum_{1\leq i_1<\ldots < i_k\leq p}\!\!\!\!\!
\bar{\mathfrak Z}_{i_1}(A)\bar{\mathfrak Z}_{i_2}(At^{-1})\bar{\mathfrak Z}_{i_3}(At^{-2})\ldots
\bar{\mathfrak Z}_{i_k}(At^{-k+1})
}
\label{mainPa}
\ee
with
\be
\!\!\!\!\!\!\!\!
\bar{\mathfrak Z}_i(At^{-s}) = \frac{\left(1-A^2t^{-4(p-i)-2-2s}\right)\left(q^{-2}-A^2t^{-2s}\right)}
{(A^2/tq)\cdot t^{-2(p-i+s)}} = \frac{\left(1+{\bf a}^2{\bf t}{\bf q}^{-4(p-i)-2-2s}\right)
\left(1+{\bf a}^2{\bf t}^3{\bf q}^{2(1-s)}\right)}
{ {\bf a}^2\cdot {\bf t}^2{\bf q}^{-2(p-i+s)}}
\ee
which also possesses the positivity property in the bold variables (\ref{newv}).
For illustration we provide just a few simplest examples of
superpolynomials for the figure eight knot:

 {\footnotesize
\be
\frac{^*{\cal P}^{4_1}_{[1]}(A|q,t)}
{ ^*\!M_{[1]}(A|\,q,t)}\
= \ 1+\{Aq\}\{At^{-1}\} =
1+{\bf t^2a^2}+{\bf q^{-2}t^{-1}}+{\bf
q^2t}+{\bf t^{-2}a^{-2}} \\
\frac{^*{\cal P}^{4_1}_{[2]}(A|q,t)}
{ ^*\!M_{[2]}(A|\,q,t)}\ = 1+\{Aq\}\{At^{-1}\}+\{Aq^3\}\{At^{-1}\}+\{Aq^3\}\{At^{-1}\}\{Aq^2\}\{Aqt^{-1}\}=
\\
={\bf a}^4{\bf q}^4{\bf t}^8+{\bf a}^2({\bf q}^{-2}{\bf t}+{\bf t}^2+{\bf t}^3+{\bf q}^2{\bf t}^4+{\bf q}^4{\bf t}^5+
{\bf q}^6{\bf t}^7)+{\bf q}^6{\bf t}^4+{\bf q}^4{\bf t}^3
+{\bf q}^2{\bf t}^2+{\bf q}^2{\bf t}+3+{\bf q}^{-2}{\bf t}^{-1}+{\bf q}^{-2}{\bf t}^{-2}+\\+
{\bf q}^{-4}{\bf t}^{-3}+{\bf q}^{-6}{\bf t}^{-4}+
{\bf a}^{-2}({\bf q}^2{\bf t}^{-1}+{\bf t}^{-2}+{\bf t}^{-3}+{\bf q}^{-2}{\bf t}^{-4}+
{\bf q}^{-4}{\bf t}^{-5}+{\bf q}^{-6}{\bf t}^{-7})+{\bf a}^{-4}{\bf q}^{-4}{\bf t}^{-8}
 \\
\frac{^*{\cal P}^{4_1}_{[11]}(A|q,t)}
{ ^*\!M_{[11]}(A|\,q,t)}\ =  1+\{Aq\}\{At^{-1}\}+\{Aq\}\{At^{-3}\}+\{Aq\}\{At^{-3}\}\{At^{-2}\}\{Aqt^{-1}\}=  \\
={\bf a}^4{\bf q}^{-4}{\bf t}^4+{\bf a}^2({\bf q}^{2}{\bf t}^3+{\bf t}^2+{\bf t}^3+{\bf q}^{-2}{\bf t}^2+
{\bf q}^{-4}{\bf t}+
{\bf q}^{-6}{\bf t})+{\bf q}^{-6}{\bf t}^{-2}+{\bf q}^{-4}{\bf t}^{-1}
+{\bf q}^{-2}+{\bf q}^{-2}{\bf t}^{-1}+3+{\bf q}^{2}{\bf t}+\\+{\bf q}^{2}+
{\bf q}^{4}{\bf t}+{\bf q}^{6}{\bf t}^{2}+
{\bf a}^{-2}({\bf q}^{-2}{\bf t}^{-3}+{\bf t}^{-2}+{\bf t}^{-3}+{\bf q}^{2}{\bf t}^{-2}+
{\bf q}^{4}{\bf t}^{-1}+{\bf q}^{6}{\bf t}^{-1})+{\bf a}^{-4}{\bf q}^{4}{\bf t}^{-4}\\
\frac{^*{\cal P}^{4_1}_{[3]}(A|q,t)}
{ ^*\!M_{[3]}(A|\,q,t)}\ = {\bf a}^6{\bf q}^{12}{\bf t}^{18}+{\bf a}^4({\bf t}^7{\bf q}^2+{\bf t}^8{\bf q}^4+
{\bf t}^9{\bf q}^4+{\bf t}^{10}{\bf q}^6+{\bf t}^{11}{\bf q}^6+{\bf t}^{12}{\bf q}^8+
{\bf t}^{13}{\bf q}^{10}+{\bf t}^{15}{\bf q}^{12}+{\bf t}^{17}{\bf q}^{14})+\\
+{\bf a}^2({\bf t}^{-2}{\bf q}^{-6}+{\bf t}^{-1}{\bf q}^{-4}+{\bf q}^{-4}+2{\bf t}{\bf q}^{-2}+
{\bf t}^2{\bf q}^{-2}+2{\bf t}^2+2{\bf t}^3+3{\bf t}^4{\bf q}^2+{\bf t}^5{\bf q}^2+
{\bf t}^5{\bf q}^4+4{\bf t}^6{\bf q}^4+
2{\bf t}^7{\bf q}^6+2{\bf t}^8{\bf q}^6+
2{\bf t}^9{\bf q}^8+\\+
{\bf t}^{10}{\bf q}^8+
{\bf t}^{10}{\bf q}^{10}+{\bf t}^{11}{\bf q}^{10}+{\bf t}^{12}{\bf q}^{12}+{\bf t}^{14}{\bf q}^{14})+
{\bf t}^{-9}{\bf q}^{-12}+{\bf t}^{-8}{\bf q}^{-10}+{\bf t}^{-7}{\bf q}^{-8}+{\bf t}^{-6}{\bf q}^{-8}+
3{\bf t}^{-5}{\bf q}^{-6}+{\bf t}^{-4}{\bf q}^{-6}+{\bf t}^{-4}{\bf q}^{-4}+\\
+4{\bf t}^{-3}{\bf q}^{-4}+
3{\bf t}^{-2}{\bf q}^{-2}+3{\bf t}^{-1}{\bf q}^{-2}+{\bf t}^{-1}+5+{\bf t}+3{\bf t}{\bf q}^2+
3{\bf t}^2{\bf q}^2+4{\bf t}^3{\bf q}^4+{\bf t}^4{\bf q}^4+{\bf t}^4{\bf q}^6+3{\bf t}^5{\bf q}^6+
{\bf t}^6{\bf q}^8+
{\bf t}^7{\bf q}^8+{\bf t}^8{\bf q}^{10}+\\
+{\bf t}^9{\bf q}^{12}
+{\bf a}^{-2}({\bf t}^{-14}{\bf q}^{-14}+{\bf t}^{-12}{\bf q}^{-12}+{\bf t}^{-11}{\bf q}^{-10}+
{\bf t}^{-10}{\bf q}^{-10}+{\bf t}^{-10}{\bf q}^{-8}+2{\bf t}^{-9}{\bf q}^{-8}+2{\bf t}^{-8}{\bf q}^{-6}+
2{\bf t}^{-7}{\bf q}^{-6}+
4{\bf t}^{-6}{\bf q}^{-4}+\\
+{\bf t}^{-5}{\bf q}^{-4}+{\bf t}^{-5}{\bf q}^{-2}+3{\bf t}^{-4}{\bf q}^{-2}+
2{\bf t}^{-3}+2{\bf t}^{-2}+{\bf t}^{-2}{\bf q}^2+2{\bf t}^{-1}{\bf q}^2+{\bf q}^4+{\bf t}{\bf q}^4+
{\bf t}^2{\bf q}^6)+\\+
{\bf a}^{-4}({\bf t}^{-17}{\bf q}^{-14}+{\bf t}^{-15}{\bf q}^{-12}+{\bf t}^{-13}{\bf q}^{-10}+
{\bf t}^{-12}{\bf q}^{-8}+{\bf t}^{-11}{\bf q}^{-6}+{\bf t}^{-10}{\bf q}^{-6}+{\bf t}^{-9}{\bf q}^{-4}+
{\bf t}^{-8}{\bf q}^{-4})+{\bf t}^{-7}{\bf q}^{-2})+{\bf a}^{-6}{\bf t}^{-18}{\bf q}^{-12}
\\
\ldots \nn \\
\\
\ee
}
\!\!\!\!\!
Note that these formulas for the superpolynomials, though natural,
are not as well tested as those for the HOMFLY polynomials (\ref{main}),
simply because there is yet nothing to compare with.
Still, they automatically include the famous formula \cite{DGR} for
the superpolynomial in the fundamental representation $R=[1]$
and also seem to reproduce the recent answers \cite{GMS} for the two-box cases,
$R=[2]$ and $R=[11]$ (they are related by the symmetry transform (\ref{symsp}).\footnote{
In fact, in \cite{GMS} a different grading was used, still it looks
plausible that those formulas are consistent with ours.}
This is, however, a very small set of evidence as compared to the list
in s.\ref{favor} for HOMFLY polynomials, and further tests are very essential.

\section{Difference equations for HOMFLY and superpolynomials}

The colored Jones and refined Jones polynomials in representation $R=\rp$
are known to satisfy {\it linear} difference equations in the variable $p$
\cite{Apol,FGS},\footnote{An even more interesting question is if the
non-linear equations also exist, generalizing {\it bi}-linear integrable Hirota-like
equations. In this direction only some preliminary results for the torus knots are available,
see \cite{MMMkn1}.}
and it is natural to ask, if such equations exist for our superpolynomials (\ref{mainP}).
In fact, one expects a whole set of such equations to exist,
a substitute of the set of the Virasoro constraints in matrix models \cite{UFN3}.
This set defines the (quantum) spectral curve
and serves as the starting point for the AMM/EO topological recursion \cite{AMM/EO}.
So far most results in this direction are obtained for the colored Jones (super)polynomials,
in particular, the recursion for the figure eight knot $4_1$ is studied in \cite{DiFu}
basing on ${\cal A}$-polynomial.
The first equation of this kind for superpolynomials is just written in
\cite{FGS} for the case of the trefoil $3_1$ (in fact, for the whole series of torus knots $(2,2k+1)$),
where it is derived from explicit
expressions for the colored superpolynomials obtained by the "evolution" method of \cite{DMMSS}.

It is, therefore, natural to do the same for our explicit expressions (\ref{mainP}).
A new possibility here is to study the equations for arbitrary $A$.
It turns out that the most natural linear equation is somewhat different from
what one always considers: they involve variation not only in $p$, but also in $A$.
Such a "basic" equation is of the first order, while for $A=q^N$ with fixed $N$
one usually gets an $N$th-order difference equation in $p$, which should be a corollary
of the "basic" one.
In the case of the superpolynomials it is also natural to expect that two "basic" equations
should exist, associated with the horizontal and vertical shifts along the Young diagram $R$.
However, only one is immediately seen from our formulas, because they are written only
for the single-line diagrams $R=\rp$.\footnote{Of course, another equation, in $t$-direction,
immediately follows for the set $R=[1^p\,]$ in (\ref{mainPa}), but in order to see both equations
at once, one should look at the answer for generic $R$.}
To write down the equation, we switch to still another notation, which makes explicit the
dependence on $p$.
Namely, instead of ${\mathfrak Z}_i(A)=\{Aq^{2(p-i)+1}\}\{At^{-1}\}$ we introduce
${\mathfrak Z}_{I|J}=\{Aq^{I}\}\{At^{-J}\}$, so that ${\mathfrak Z}_i = {\mathfrak Z}_{2p-2i+1|1}$.
We also need a condensed notation for the action of the dilatation operator $\hat D_q:\ A\rightarrow qA$,
\be
{\mathfrak Z}_{I|J}^{(s)}(A) = \hat D_q^s {\mathfrak Z}_{I|J}(A) =
{\mathfrak Z}_{I|J}(q^sA)=\{Aq^{I+s}\}\{Aq^st^{-J}\}
\ee
Then, the first few normalized superpolynomials
$P_{\rp} = \frac{^*{\cal P}^{4_1}_{\rp}(A|q,t)}{^*\!M_{\rp}(A|q,t)}$  in (\ref{mainP}) are:
\be
P_{[0]} = 1,\\
P_{[1]} = 1 + {\mathfrak Z}_{1|1}, \\
P_{[2]} = 1  + \Big({\mathfrak Z}_{3|1} + {\mathfrak Z}_{1|1}\Big)
+ {\mathfrak Z}_{3|1}{\mathfrak Z}_{1|1}^{(1)},\\
P_{[3]} = 1  + \Big({\mathfrak Z}_{5|1} + {\mathfrak Z}_{3|1} + {\mathfrak Z}_{1|1}\Big)
+ \Big({\mathfrak Z}_{5|1}{\mathfrak Z}_{3|1}^{(1)}+ {\mathfrak Z}_{5|1}{\mathfrak Z}_{1|1}^{(1)}
+ {\mathfrak Z}_{3|1}{\mathfrak Z}_{1|1}^{(1)}\Big)
+ {\mathfrak Z}_{5|1}{\mathfrak Z}_{3|1}^{(1)}{\mathfrak Z}_{1|1}^{(2)},\nn\\
\ldots
\ee
so that
\be
P_{[1]}-P_{[0]} = {\mathfrak Z}_{1|1} = {\mathfrak Z}_{1|1} P_{[0]}, \\
P_{[2]}-P_{[1]} = {\mathfrak Z}_{3|1}\Big(1+{\mathfrak Z}_{1|1}^{(1)}\Big) = {\mathfrak Z}_{3|1} P_{[1]}^{(1)}, \\
P_{[3]}-P_{[2]} = {\mathfrak Z}_{5|1}\Big(1+{\mathfrak Z}_{1|1}^{(1)}+{\mathfrak Z}_{3|1}^{(1)}
+ {\mathfrak Z}_{3|1}^{(1)}{\mathfrak Z}_{1|1}^{(2)}\Big) = {\mathfrak Z}_{5|1} P_{[2]}^{(1)}, \nn \\
\ldots
\ee
and in general
\be
\boxed{
P_{[p+1]}(A)-P_{\rp}(A) = {\mathfrak Z}_{2p+1|1}(A)P_{\rp}^{(1)}(A)
= \{Aq^{2p+1}\}\{At^{-1}\}P_{\rp}(qA)
}
\ee
Here
$t$ is just a parameter,
and one obtains an equation for the normalized HOMFLY polynomials just by putting $t=q$:
\be
\boxed{\
h(A,p+1)-h(A,p)-
\left(A^2q^{2p}+{1\over A^2q^{2p}}-q^{2p+2}-{1\over q^{2p+2}}\right)h(qA,p)=0\
}
\ee
for $h(A,p) = {^*{\cal H}}^{4_1}_{\rp}(A)/\SS_{\rp}(A)$.
This equation can be rewritten
in terms of "the quantum ${\cal A}$-polynomial",
\be
\hat {\cal A}\left(\frac{^*{\cal H}^{4_1}_{\rp }(A|\,q)}{\SS_{\rp}(A|\,q)}\right) =0,\ \ \ \ \ \ \ \  \ \ \
\
\hat {\cal A}\equiv \hat l-1-\left(A^2\hat m^2+{1\over A^2\hat m^2}-q^{2}\hat
m^2-{1\over q^{2}\hat m^2}\right)\hat D
\ee
where, as usual, $\hat l f(A,p)=f(A,p+1)$
and $\hat m f(A,p)=q^p f(A,p)$.
As already mentioned, this equation differs from the standard Jones polynomial
equation as it involves two variables: $p$ (the representation) and $A$ (the
group), on the other hand, it
is the first order difference equation, while the Jones polynomial equation is
the second order difference equation
in one variable $p$. One may expect difference equations of higher orders in
this one variable in the case of $A=q^N$
with $N>2$, however, for generic $A$ it is unclear if an equation of such a
type exists at all.

\section{On possible generalizations}

There are two obvious directions to further
generalize and explore the result (\ref{main}):
\begin{itemize}
\item from $R=\rp $ to arbitrary $R$ (arbitrary Young diagram
$R = \{p_1\geq p_2\geq \ldots \geq 0\}$),
\item  from ${\cal K} = (1,-1|1,-1)$
to entire series of 3-strand knots
${\cal K} = (1,-1)^n = (1,-1|1,-1|\ldots |1,-1)$, a simple generalization of
the torus knot family ${\cal K} = [3,n] = (1,1)^n$ (these are knots for
$n$, indivisible by $m=3$ and $3$-component links otherwise),
\item one can also convert to the character decomposed form of the formula
and find for this family an analogue of the  Rosso-Jones formula \cite{chi}
for the torus knots (this was done in \cite{MMMkn2} for the fundamental
representation only, so there was no chance of formulating a counterpart
of the Adams rule).
\end{itemize}


\noindent
As concerns the first item in this list,
eqs.(\ref{main}) and (\ref{mainas}) are expected to extended naturally to a
full formula.
Like in the case of superpolynomials, more suited for generalization
is not (\ref{main}) itself, but its reformulation (\ref{amain}), which
is just a sum of contributions from al subsets of boxes in the Young diagram $R$
taken with unit weights.
However, though easily conjectured, the formulas for arbitrary $R$
can not be sufficiently tested at the moment due to the lack of any alternative
calculations.
To avoid mixing reliable formulas with speculations,
we discuss  generic representations elsewhere.

\section*{Acknowledgements}

Our work is partly supported by Ministry of Education and Science of
the Russian Federation under contract 14.740.11.0347, by NSh-3349.2012.2,
by RFBR grants 10-02-00509 (A.Mir.), 10-02-00499 (A.Mor.), 10-02-01315
(And.Mor.) and
by joint grants 11-02-90453-Ukr, 12-02-91000-ANF,
11-01-92612-Royal Society.
The research of H.~I.~ and A.Mir.
is supported in part by the Grant-in-Aid for Sci\-en\-tific Research
(23540316)
from the Ministry of Education, Science and Culture, Japan, and that of A.Mor.
by
by JSPS Invitation Fellowship Program for Research in Japan (S-11137).
Support from JSPS/RFBR bilateral collaboration "Synthesis of integrabilities
 arising from gauge-string duality" (FY2010-2011: 12-02-92108-Yaf-a) is
gratefully appreciated.

\end{document}